\documentclass[
 amsmath,amssymb,
 reprint,%
]{revtex4}

\usepackage{graphicx}
\usepackage{dcolumn}
\usepackage{bm}

\usepackage[utf8]{inputenc}
\usepackage[T1]{fontenc}
\usepackage{mathptmx}
\usepackage{float}
\usepackage{color}
\newcommand{\ks}{\textcolor{black}}

\usepackage{color}

\def\sin{{\rm{sin}}}

\def\u{{\rm{\bf u}}}

\def\f{{{\rm{\bf f}}}}

\begin{document}

\title{\ks{Simulations of a weakly conducting droplet under the influence of an alternating electric field}}

\author{Kirti Chandra Sahu}
\email{ksahu@che.iith.ac.in}
 \homepage{http://www.iith.ac.in/~ksahu.}
\affiliation{%
Department of Chemical Engineering, Indian Institute of Technology Hyderabad, Sangareddy 502 285, Telangana, India}%
\author{Manoj Kumar Tripathi}
\affiliation{Indian Institute of Science Education and Research Bhopal 462 066, Madhya Pradesh, India}
 \author{Jay Chaudhari}
\affiliation{Indian Institute of Science Education and Research Bhopal 462 066, Madhya Pradesh, India}
\author{Suman Chakraborty}
\affiliation{Department of Mechanical Engineering, Indian Institute of Technology Kharagpur, Kharagpur 721302, India}

\date{\today}


\begin{abstract}
\ks{We investigate the electrohydrodynamics of an initially spherical droplet under the influence of an external alternating electric field by conducting axisymmetric numerical simulations using a charge-conservative volume-of-fluid based finite volume flow solver. The mean amplitude of shape oscillations of a droplet subjected to an alternating electric field for leaky dielectric fluids is the same as the steady-state deformation under an equivalent root mean squared direct electric field for all possible electrical conductivity ratio $(K_r)$ and permittivity ratio $(S)$ of the droplet to the surrounding fluid. In contrast, our simulations for weakly conducting media show that this equivalence between alternating and direct electric fields does not hold for $K_r \ne S$. Moreover, for a range of parameters, the deformation obtained using the alternating and direct electric fields is qualitatively different, i.e. for low $K_r$ and high $S$, the droplet becomes prolate under alternating electric field but deforms to an oblate shape in the case of the equivalent direct electric field. A parametric study is conducted by varying the time period of the applied alternating electric field, the permittivity and the electrical conductivity ratios. It is observed that while increasing $K_r$ has a negligible effect on the deformation dynamics of the droplet for $K_r<S$, it enhances the deformation of the droplet when $K_r>S$ for both alternating and direct electric fields. We believe that our results may be of immense consequence in explaining the morphological evolution of droplets in a plethora of scenarios ranging from nature to biology.}
\end{abstract}

\maketitle

\section*{1~~Introduction}
\vspace{-4mm}
\label{sec:intro}
Electrically driven dynamics of a liquid droplet suspended in another medium has been a subject of intense research from several decades due to its relevance in industrial applications \cite{basaran2013nonstandard}, microfluidics \cite{stone2004engineering,nath2018cross,santra2018electrohydrodynamics,mandal2017effect}, biological systems \cite{vlahovska2018electrohydrodynamics} and natural phenomena such as electrification of rain, raindrops bursting in thunderstorms and electrification of the atmosphere \cite{simpson1909xv,whybrew1952electrification,o1953distortion}. Significant research efforts, therefore, have been directed to address various facets of the coupling between electromechanics and hydrodynamics over various spatial and temporal scales \cite{melcher1969electrohydrodynamics}. Despite a phenomenal advancement in the field over the years, however, there remain many deficits in developing a generalised theoretical understanding of the parities and disparities of the dynamical response of a droplet under alternating (AC) and direct (DC) electrical fields. \ks{This deficit stems from the complexities in capturing the underlying physics as well as the assumptions involved regarding the electrohydrodynamic properties of the fluids.}

Reported research has revealed that the main electrical parameters that govern the shape evolution of a droplet under direct or alternating electric field are the electrical conductivity ratio and the permittivity ratio between the droplet and surrounding fluid. The electrohydrodynamics (EHD) of a droplet under the action of an applied direct electric field have been investigated by several researchers in the past considering perfect \citep{sherwood1988,hua2008a,lin2012} and leaky dielectric media \citep{tomar2007,nath2018cross}. While the underlying physics under {DC field} have been addressed in details from deep-rooted theoretical \cite{taylor1966studies}, experimental \cite{torza1971} and numerical \cite{sherwood1988,hua2008a,lin2012,tomar2007,mhatre2013drop,nath2018cross,lanauze2015nonlinear,das2017nonlinear,das2017electrohydrodynamics} considerations, there is a compelling need to assess a possible straight forward extrapolation of identical inferences for {alternating electric field} as well \cite{soni2018electrohydrodynamics}. In a recent study, Esmaeeli and Halim \cite{esmaeeli2018electrohydrodynamics} provided extensive two-dimensional numerical simulations for \ks{leaky dielectric systems}, to predict that a droplet in an {alternating} electric field undergoes shape oscillations about the steady-state deformation observed under a root-mean-squared equivalent DC field, for all possible electrical conductivity ($K_r$) and permittivity $(S)$ ratios. \ks{However, this is not true for general class of fluids as reported in the experimental study of Torza {\it et al.} \cite{torza1971}. This motivates us to re-examine this problem without the leaky dielectric assumption. We investigate the electrohydrodynamics of an initially spherical droplet under the influence of an external alternating electric field by conducting axisymmetric numerical simulations using a charge-conservative volume-of-fluid based finite volume flow solver without the leaky dielectric assumption. In order to isolate the effect of the electric field, it is assumed that the droplet and surrounding medium have the same density.  The dynamic viscosity of the fluids are also assumed to be the same. We found that the equivalence between the deformation observed under DC and AC fields exists for all values of $K_r$ and $S$ in the leaky dielectric limit, as reported by the previous studies \cite{esmaeeli2018electrohydrodynamics}. For weakly conducting systems, our results reveal that under the AC field, the droplet becomes prolate (elongates in the direction of the electric field), whereas in the case of the equivalent DC electric field, the droplet becomes oblate for high $S$ and low $K_r$ values. For some range of parameters, we found that the deformation behaviour (oblate/prolate) predicted by the direct electric field theory \cite{melcher1969electrohydrodynamics} is different from that obtained by the simulations with an equivalent alternating electric field.}

\section*{2~~Formulation}
\label{sec:num}

\vspace{-4mm}

\begin{figure}
\centering
\includegraphics[width=0.25\textwidth]{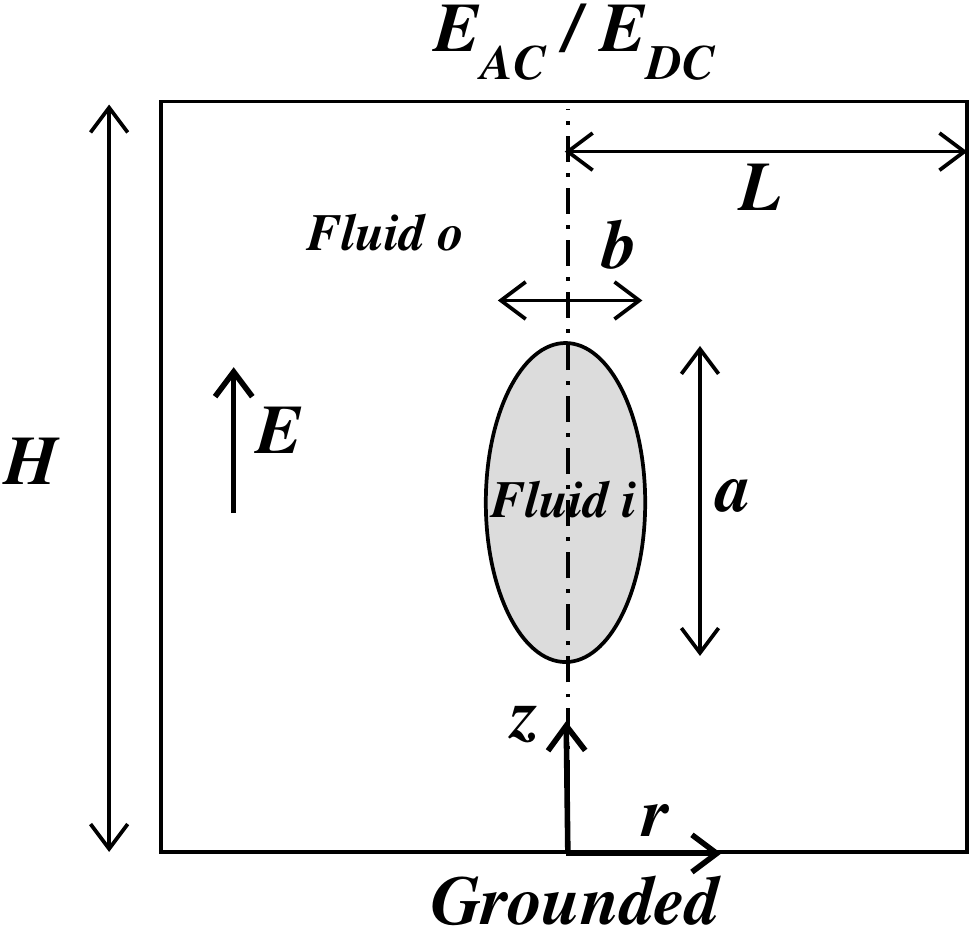} 
\caption{Schematic diagram (not to scale) of a droplet (fluid `$i$') in another immiscible fluid (fluid `$o$') under an {alternating} electric field $(E_{AC})$ or {direct} electric field $(E_{DC})$ applied in the axial direction.} 
\label{fig1}
\end{figure}

The dynamics of an initially spherical droplet of radius, $R$ inside another immiscible fluid under the influence of an external electric field is investigated via axisymmetric numerical simulations.  \ks{The schematic diagram presented in Fig. \ref{fig1} depicts an axisymmetric computational domain of size $(H \times L)=(8 R \times 8R)$, with the droplet center at $(0,4R)$.} The droplet (fluid `$i$') and the surrounding liquid (fluid `$o$') are assumed to be incompressible and Newtonian. \ks{In order to isolate the effect of electric field from other forces (such as gravity) on the droplet dynamics, the fluids are also assumed to have the same density, $\rho$. The viscosities ($\mu$) of the fluids are also assumed to be the same.} The interfacial surface tension is denoted by $\gamma$. An axisymmetric cylindrical coordinate system $(r,z)$ is used to model the flow dynamics. \ks{The electric field is applied at the top wall ($z=H$) by connecting it to an alternating power source and the bottom wall at $z=0$ is grounded. The alternating electric field is given by
\begin{equation}
E_{AC} (t) \equiv {\psi_{AC} (t) \over H} = {\psi_0 \over H}  \sin \left({2 \pi t \over T_p} \right),\label{EAC}
\end{equation} 
where, $t$ represents time, and $E_{AC} (t)$, $T_p$, $\psi_{AC} (t)$ and $\psi_0$ are the electric field, the time period of the {alternating} electric forcing, the electric potential and its magnitude, respectively.}

The droplet deformation in the presence of the AC field is compared against that observed in an equivalent DC field $(E_{DC}= {\psi_{DC} / H}$) with its magnitude equal to the root-mean-squared (rms) magnitude of the {alternating} electric field, such that
\begin{eqnarray}
E_{DC} &=& \lim_{t \to\infty}  \sqrt{{1 \over 2 T_p}  \int_{-T_p}^{T_p} \left (E_0 \sin \left({2 \pi t \over T_p} \right) \right)^2 dt}  = {E_0 \over \sqrt{2}},
\label{equivDC}
\end{eqnarray}
wherein $E_0=(\psi_0/ H)$ and $\psi_{DC}$ is the electric potential of the equivalent DC field. 

Under the influence of electric field the initially spherical droplet deforms to an ellipsoid shape with $a$ and $b$ as the length and breadth of the droplet in the directions parallel and perpendicular to the applied electric field. Thus, the degree of deformation, $D$ can be defined as ${(a - b) / (a + b)}$, such that $D>0$ and $D<0$ correspond to a prolate shape and an oblate shape, respectively; $D=0$ represents a spherical droplet. 
\ks{
\subsubsection*{2.1~~Governing equations}
\vspace{-4mm}
Under the influence of an electric field the droplet dynamics is governed by the continuity and the Navier-Stokes equations with an additional body force term associated with the electric field. The dimensional governing equations are given by 
\begin{eqnarray}
\nabla \cdot \u &=& 0, \label{contd} \\
\rho \Big [ {\partial \u \over \partial t} + \u \cdot \nabla \u \Big ] &=& -\nabla p + \nabla \cdot \left [\mu (\nabla \u + \nabla \u^T) \right]  + \delta \gamma \kappa {\bf n} + \f_e. ~~~~~\label{NSd}
\end{eqnarray}
Here, ${\bf u}=(u,w)$ is the velocity field, wherein $u$ and $w$ represent the velocity components in the $r$ and $z$ directions, respectively; $p$ is the pressure field; $\delta$ is a Dirac-delta function which is zero everywhere except at the interface; ${\bf n}$ is the unit normal to the interface pointing towards fluid `o'. The body force term for the applied electric field, $\f_e$ is given by
\begin{equation}
\f_e = \nabla \cdot {\cal M} = \rho_e E - \frac{1}{2}  E^2 \nabla \epsilon,
\label{eqfe}
\end{equation}
where ${\cal M} = \epsilon \left [E \otimes E - \frac{1}{2} \left (E \cdot E \right ) I \right ]$ is the Maxwell's stress tensor, wherein $I$ is the appropriate identity tensor, $\epsilon$ represents the electric permittivity and $E$ is the applied electric field strength (AC/DC). In the absence of magnetic field, the electric field can be assumed to be irrotational, i.e. $\nabla \times E = 0$, and the electric field can be expressed in terms of the electric potential ($\psi$), such that $E = -\nabla \psi$. Thus, the Gauss law of the volumetric free charge density ($\rho_e$) is given by 
\begin{equation}
\nabla \cdot \left ( \epsilon \nabla \psi \right) = -\rho_e.
\end{equation}
The free charge density around a fluid particle decays with a time scale, $t_e = \epsilon/K$ (known as electric relaxation time), where $K$ is the electrical conductivity. The viscous time scale is given by $t_v = \rho R^2/ \mu$.  In case of a conducting fluid with $t_e << t_v$, the charge accumulates at the interface almost instantaneously, i.e the charge conservation in the bulk fluid can reach to a steady state much faster than the fluid motion. When both fluids have low electrical conductivities, $t_e >> t_v$, then the medium is known as perfect dielectric. In this case, there are no free charge carriers, i.e. $\rho_e=0$. Note that we do not make such an assumption and solve the following bulk charge conservation equation \cite{lopez2011charge}: 
\begin{equation}
{\partial \rho_e \over \partial t} + \nabla \cdot \left (\rho_e \u \right) = \nabla \cdot \left ( K \nabla \psi \right). 
\label{elect4}
\end{equation}
The permittivity $(\epsilon)$ and the electrical conductivity $(K)$ are assumed to depend on the volume fraction, $c$ of fluid `$i$' as
\begin{eqnarray}
{\epsilon}= (1-c) \epsilon_o + c \epsilon_i, ~ {K}= (1-c) K_o + c K_i,
\end{eqnarray}
where $\epsilon_i$, $\epsilon_o$ and $K_i$, $K_o$ are the electrical permittivities and conductivities of the droplet and surrounding fluid, respectively. 
\subsubsection*{2.2~~Nondimensionalisation}
The following scaling is used to non-dimensionalise the governing equations:
\begin{eqnarray}
(x,z) ={R} \left({\widetilde x,\widetilde z}\right), ~ (t, T_p) ={R / V_s} (\widetilde t, \widetilde T_p),  ~ {\bf u} = V_s\tilde{\bf u}, ~ p= \rho {V_s^2} \widetilde p,  \nonumber \\
 \epsilon = \epsilon_{o}\widetilde \epsilon,  ~ K = K_{o}\widetilde K, ~ \delta = \widetilde \delta/R, ~\rho_e = \left( {\rho {V_s^2} / E_s R } \right) \widetilde \rho_e, ~~~\nonumber \\
 E = E_s \widetilde {E} , ~ (\psi,\psi_0)= R E_s (\widetilde {\psi}, \widetilde {\psi_0}), ~~~~~~~~~~~~ \label{eq:scaling}
\end{eqnarray}
where $V_s \equiv \sqrt{\gamma /\rho R}$ is the reference velocity and $E_s \equiv \sqrt{\gamma /R \epsilon_0}$ is the characteristic electric field strength. The tildes designate dimensionless quantities. After dropping tildes from all nondimensional variables, the governing dimensionless equations are given by
\begin{eqnarray}
\nabla \cdot \u &=& 0, \label{conti}
\\
{\partial \u \over \partial t} +  \u \cdot \nabla \u  &=& -\nabla p + {1 \over Re} \nabla \cdot \left [\mu (\nabla \u + \nabla \u^T) \right]  + \delta {\nabla \cdot {\bf n}  \over We} {\bf n}  ~~~~ \nonumber \\ &+&\left( \rho_e E -  {E^2 \nabla \epsilon \over 2 {\chi}} \right), ~~~~~~~~~ \label{NS} 
\\
\nabla \cdot \left ( \epsilon \nabla \psi \right) &=& -\chi \rho_e, 
\\
E &=& -\nabla \psi.
\\
{\partial \rho_e \over \partial t} + \nabla \cdot \left (\rho_e \u \right) &=&{1 \over \chi O_c}\nabla \cdot \left ( K \nabla \psi \right).
\label{elect4b}
\end{eqnarray}
We solve the following advection equation for the volume fraction, $c$ of fluid `$i$' in order to track the interface separating the droplet $(c=1)$ and the surrounding fluid $(c=0)$:
\begin{equation}
{\partial c \over \partial t} + \u \cdot \nabla c  = 0.
\end{equation}
The dimensionless permittivity and electrical conductivity are given by
\begin{eqnarray}
\epsilon = (1-c) +  c S, ~ {K} = (1-c) + c K_r.
\label{vis:model}
\end{eqnarray}
The various dimensionless numbers are the Reynolds number $(Re (\equiv \rho V_s R /\mu))$, the electrical conductivity ratio $(K_r (\equiv K_i/K_o))$, the permittivity ratio $(S (\equiv \epsilon_i/\epsilon_o))$, and the dimensionless number associated with Ohmic charge conduction, $O_c$ is ${V_s \epsilon_o / K_o R}$. In addition to these dimensionless numbers, there are two more dimensionless numbers, namely the Weber number $(We (\equiv \rho V_s^2 {R}/\gamma))$ and the electro-gravitational number $(\chi (\equiv \rho {V_s}^2 /\epsilon_o E_s^2))$, which are equal to one due to the present choice of the scales. Note that for a leaky dielectric system, $O_cRe$, which is equivalent to the electric Reynolds number \cite{esmaeeli2018electrohydrodynamics}, is $\ll 1$.
\subsubsection*{2.3~~Boundary conditions} 
The no-slip and no-penetration conditions are imposed at the top and bottom boundaries (electrodes), the free-slip boundary condition is applied at the side boundary, and the symmetry boundary condition is imposed at the centerline of the computational domain. The temporal variations of the degree of deformation, $D$ of the droplet due to the application of an applied electric field (AC/DC) is investigated. Previously, several researchers have investigated shape oscillations of a droplet without any external electric field \cite{zhang2019short,deka2019dynamics,balla2019shape,agrawal2017nonspherical}.} 
 
\begin{figure}
\centering
\hspace{0.5cm}{(A)} \\
 \includegraphics[width=0.3\textwidth]{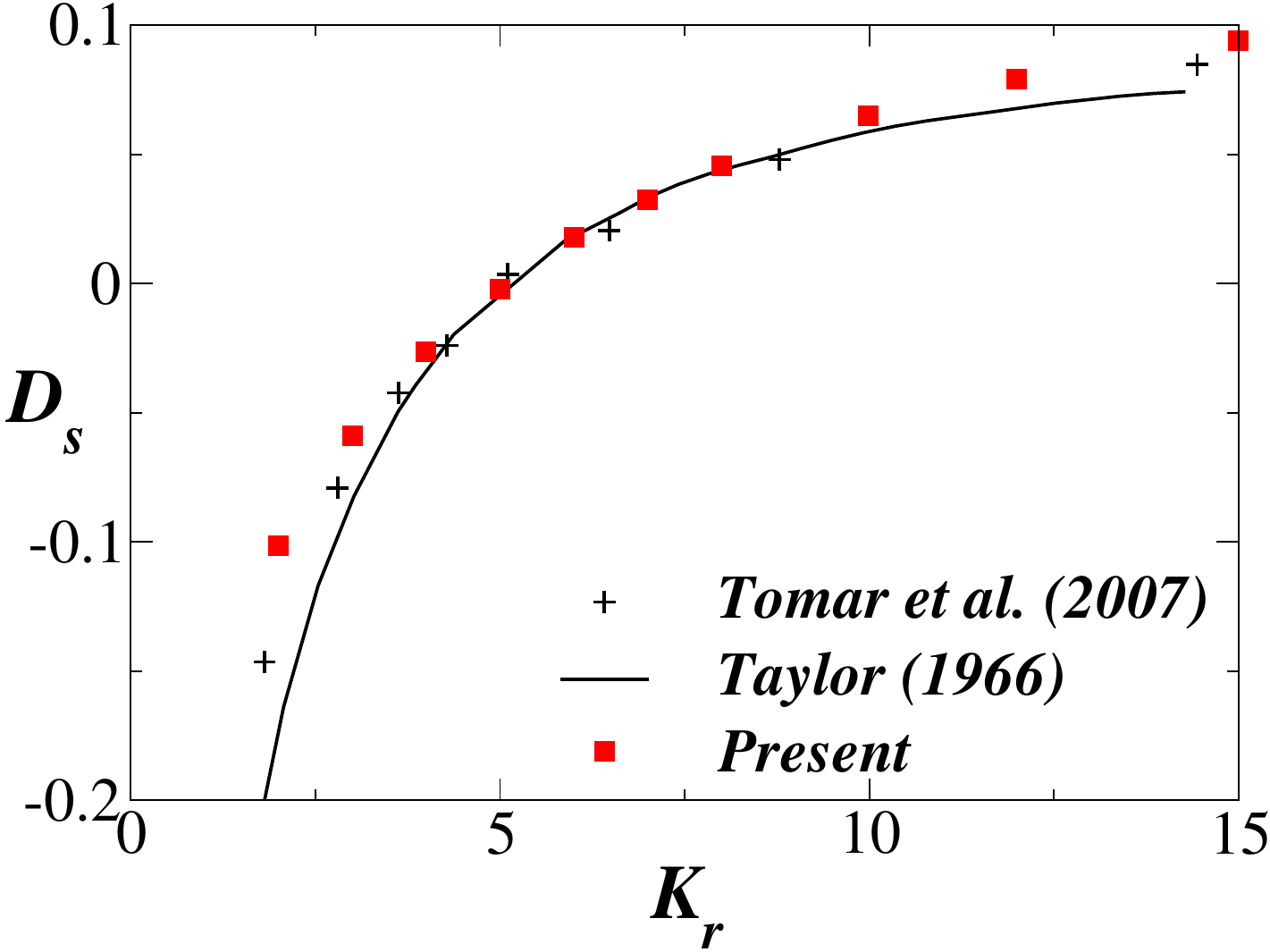} \\ 
\hspace{0.5cm}{(B)} \\
 \includegraphics[width=0.3\textwidth]{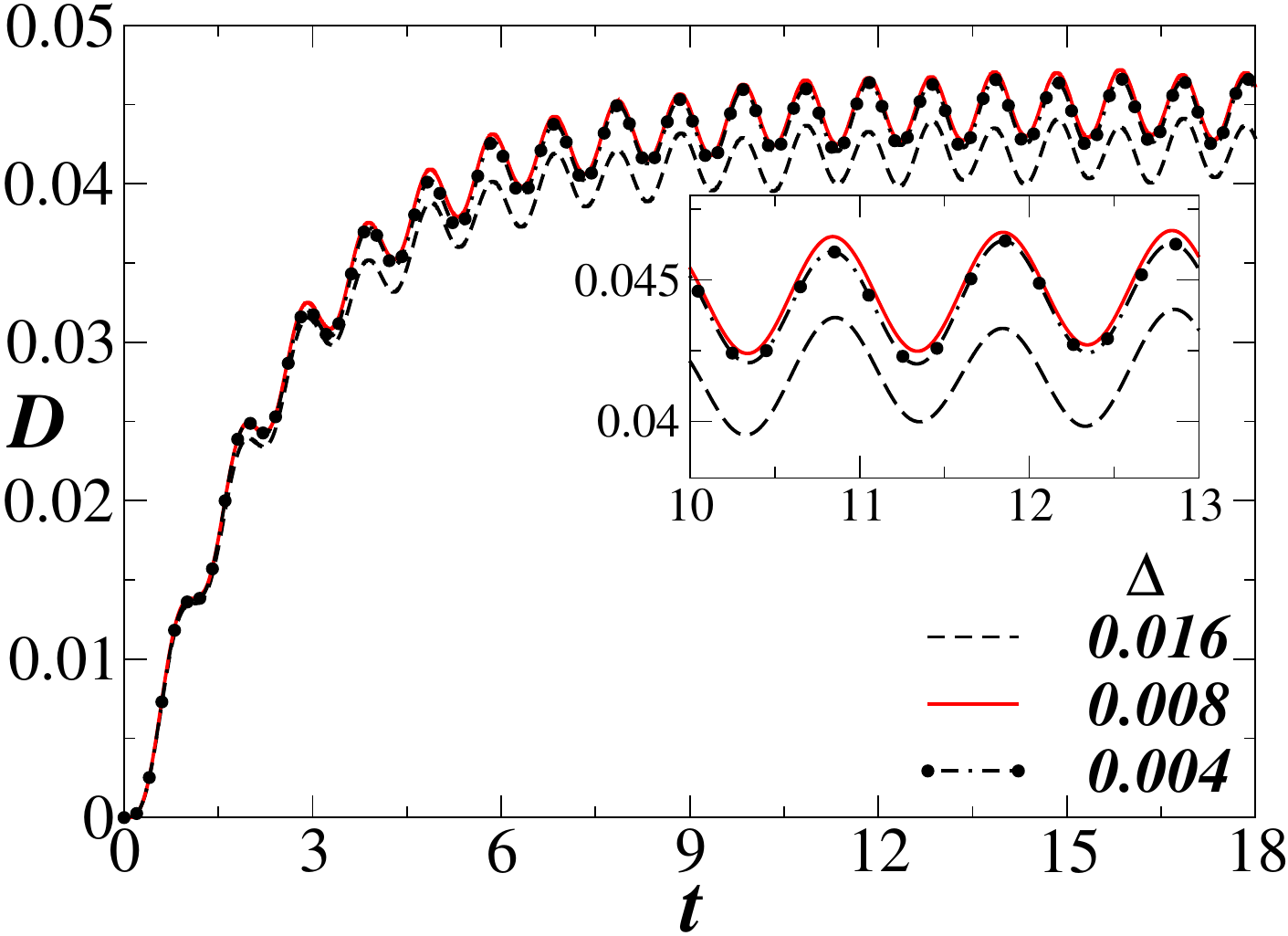} 
\caption{(A) The variations of $D_s$ with the electrical conductivity ratio, $K_r$ for $Re =0.746$, $O_c=2.68$, $E_{DC}=0.428$ and $S=10$. (B) The temporal variations of $D$ obtained using three different grids under an {alternating} electric field with $E_0=0.5$ ($\psi_0=4$) and $T_p=2$. The remaining parameter values are $Re =1$, $O_c=10$, $S=10$ and $K_r=2$. The inset is a magnified view showing the oscillations of $D$ at $10 \le t \le 13$.}
\label{fig2}
\end{figure}
\vspace{-4mm}
\section*{3~~Numerical method and validation}
\vspace{-4mm}
A volume-of-fluid (VoF) method based on a finite volume framework is used to simulate the electrohydrodynamics of a droplet in a conducting medium. An open source fluid flow solver, {\it Basilisk (http://basilisk.fr)} \cite{popinet2009} is used that employs a height-function based balanced force continuum surface force (CSF) formulation for the computation of the surface tension force. This flow solver was used in our previous studies to study the dynamics of bubbles and droplets \cite{balla2020effect,balla2020numerical,balla2019shape,balla2019non}. A charge-conservative approach is implemented by including the electric force into the Navier-Stokes equations \cite{lopez2011charge} considering both convection and conduction of the free charges. The numerical method used in the present study is similar to that of L{\'o}pez-Herrera {\it et al.} \cite{lopez2011charge}, where several validation exercises were performed. 

Further in order to validate our flow solver, the deformation of the droplet under the application of a DC electric field has been compared against the previous computational \cite{tomar2007} and theoretical \cite{taylor1966} studies in Fig. \ref{fig2}A. Here, the system is assumed to be neutrally buoyant and the simulations are performed till the steady state is reached. The parameters considered are $Re =0.746$, $O_c=2.68$, $E_{DC}=0.428$ and $S=10$. Taylor \cite{taylor1966} conducted a linearised asymptotic analysis by assuming both the fluids to be conducting and highly viscous (low $Re$), and derived an expression for $D_s$, which is given by (in the dimensionless form)
\begin{equation}
D_s =  {9 {E_{0}}^2 \over 16 (2+K_r)^2} \left [ 1 + {K_r}^2 - 2S + {3 \over 5} (K_r-S) {2 + 3 \mu_r \over 1+ \mu_r} \right], \label{theory}
\end{equation}
where $\mu_r$ is the ratio of viscosity of the droplet and the surrounding fluid, which is set to one in the present study. The numerical simulations are performed for different values of $K_r$. It can be seen in Fig. \ref{fig2}A that the droplet exhibits an oblate shape at low values of $K_r$, and upon increasing the value of $K_r$ the droplet deforms to a prolate shape via a spherical shape at an intermediate value of $K_r$. It can be seen that the present result agrees well with the previous studies. 

As the main focus of the present study is to investigate the droplet deformation dynamics under the influence of an {alternating} electric field, in Fig. \ref{fig2}B, we have conducted a grid convergence test for a typical set of parameters under the influence of an {alternating} field. The parameters considered for these simulations are $E_0=0.5$ (corresponds to $\psi_0=4$ in this case) and $T_p=2$ with the remaining parameter values being $Re =1$, $O_c=10$, $S=10$ and $K_r=2$. The simulations are performed using three meshes with dimensionless cell sizes, $\Delta=0.016$, 0.008 and 0.004. It can be seen that degree of deformation, $D$ undergoes an oscillatory variation in this case. Inspection of Fig. \ref{fig2}B reveals that we get acceptable grid-converged results for $\Delta \le 0.008$, but the result obtained using $\Delta=0.016$ under-predicts the degree of deformation of the droplet. Thus, $\Delta = 0.008$ is used to generate the rest of the results presented in this study. 

\section*{4~~Results and discussion}
\label{sec:dis}
\vspace{-4mm}
 \begin{figure}[h]
\centering
\includegraphics[width=0.3\textwidth]{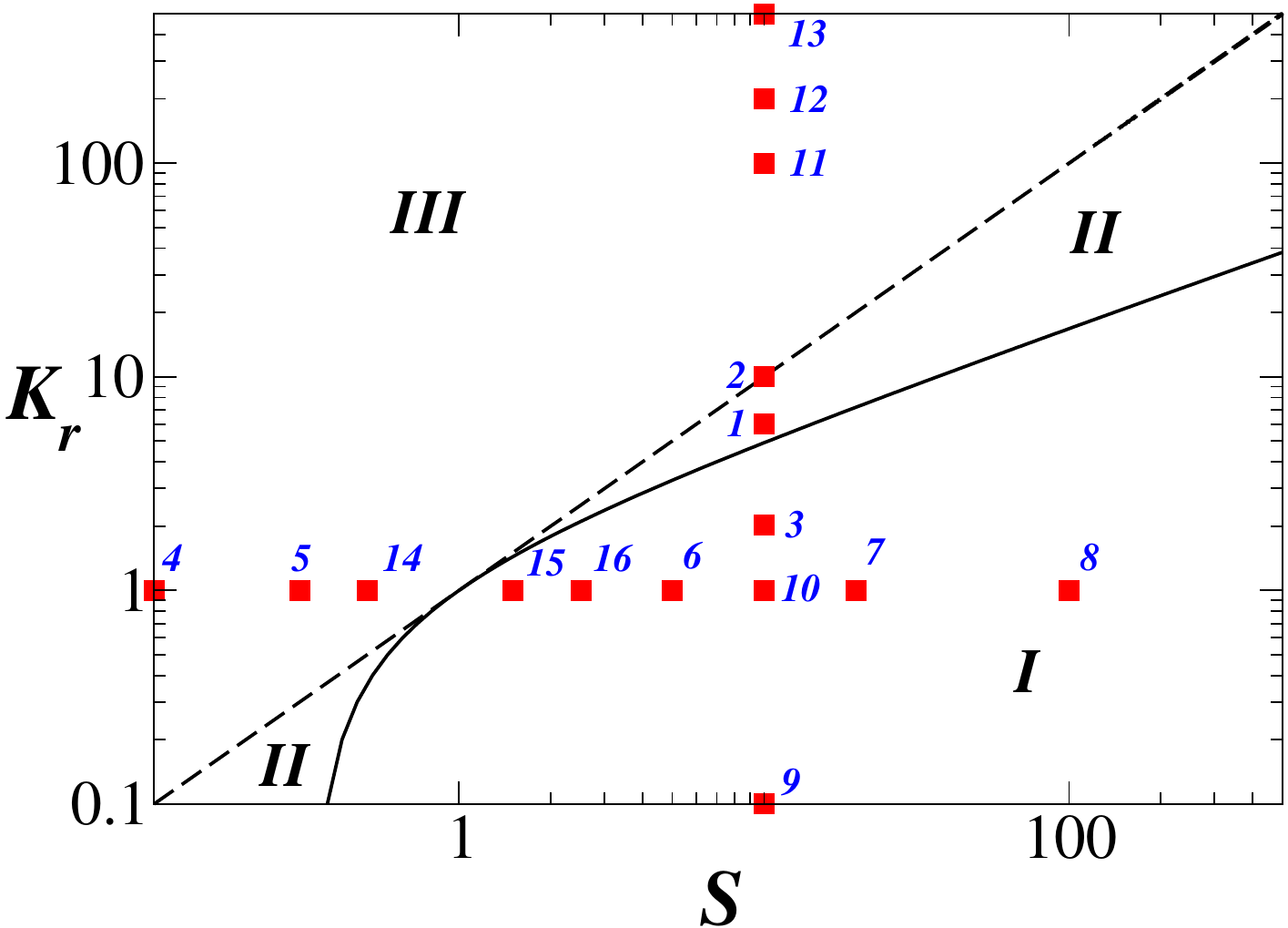}
\caption{The deformation-circulation map of a droplet under an {alternating} electric field \cite{torza1971}. The dashed and solid lines correspond to $S=K_r$ (the zero-circulation line) and $K_r^2 + K_r + 1 - 3S =0$ (the zero-deformation curve). The red filled square symbols represent the values of $S$ and $K_r$ considered in the present study. }
\label{defmap}
\end{figure}

\begin{figure}[h]
\centering
\hspace{0.2cm} {(A)} \hspace{3.8cm} {(B)}\\
\includegraphics[width=0.23\textwidth]{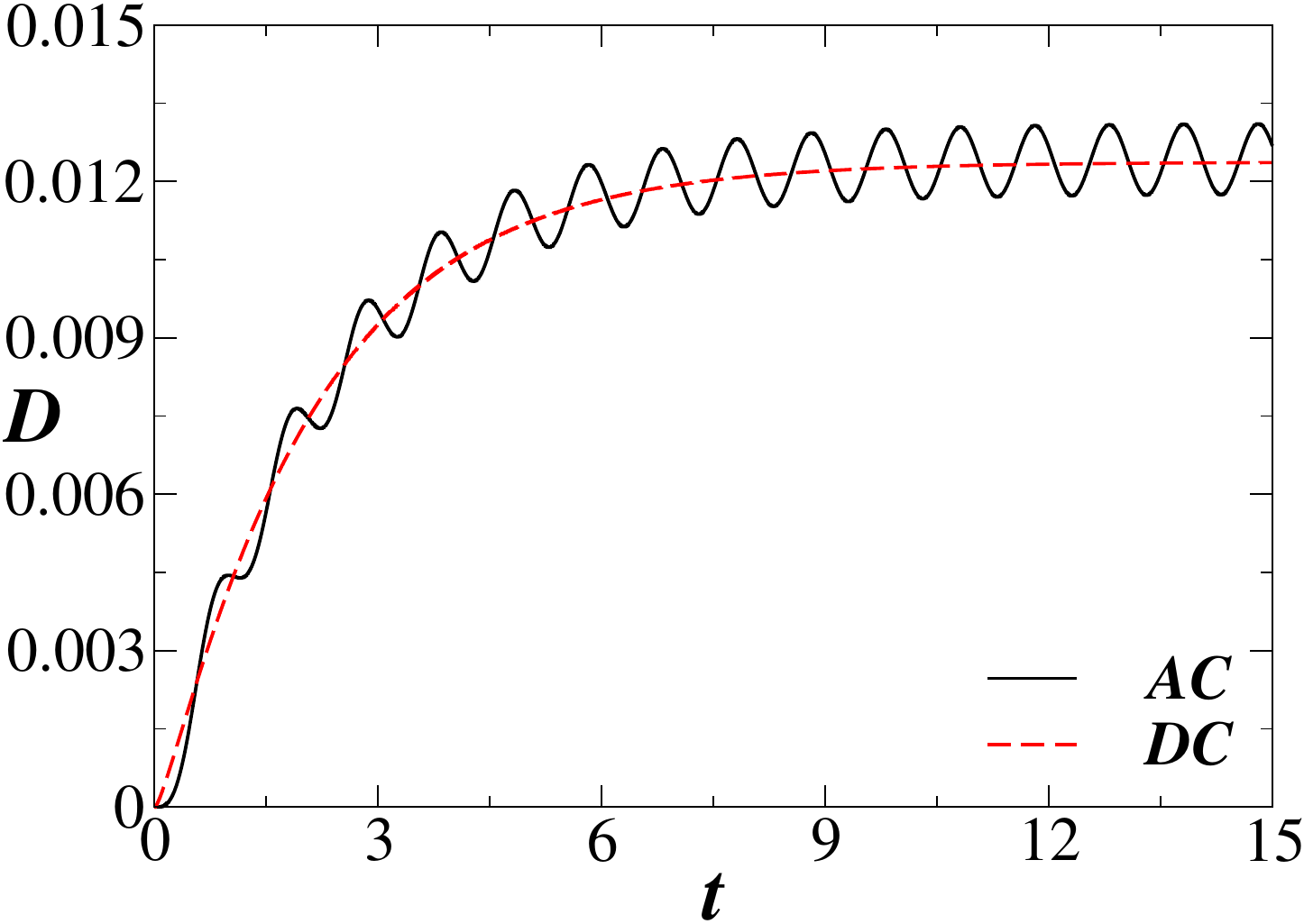}  \includegraphics[width=0.23\textwidth]{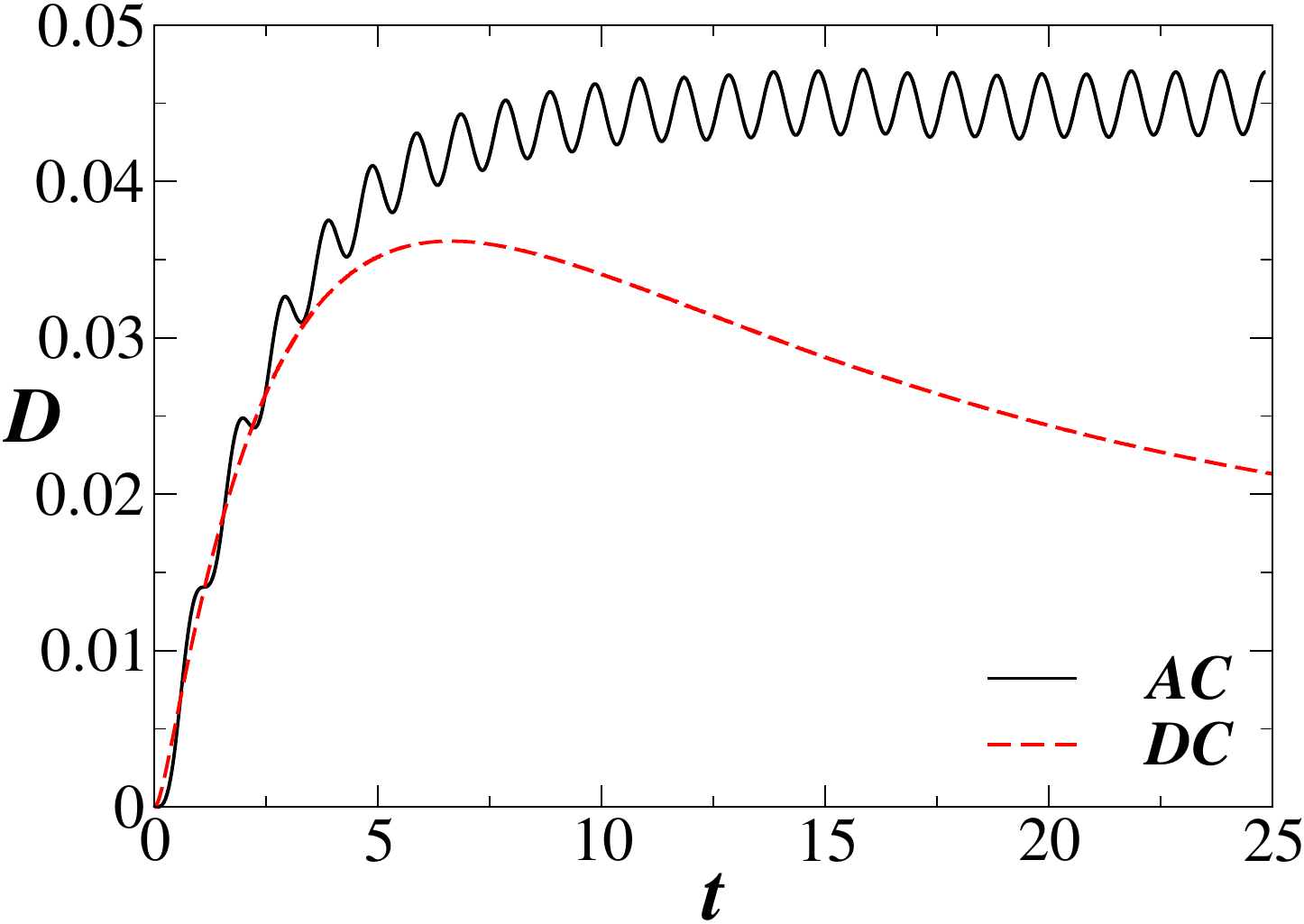}  \\
\hspace{0.2cm} {(C)} \hspace{3.8cm} {(D)}\\
\includegraphics[width=0.23\textwidth]{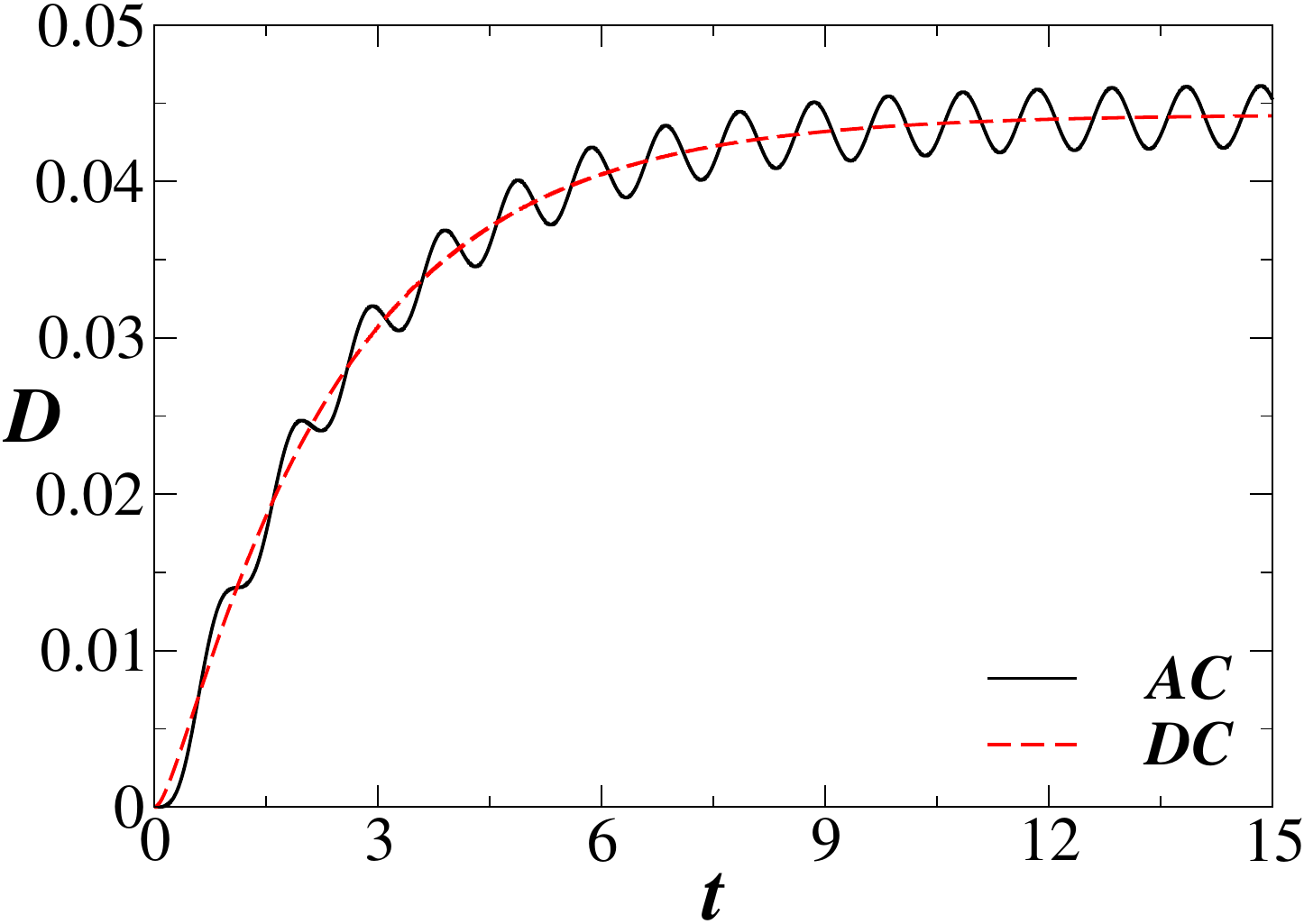} \includegraphics[width=0.23\textwidth]{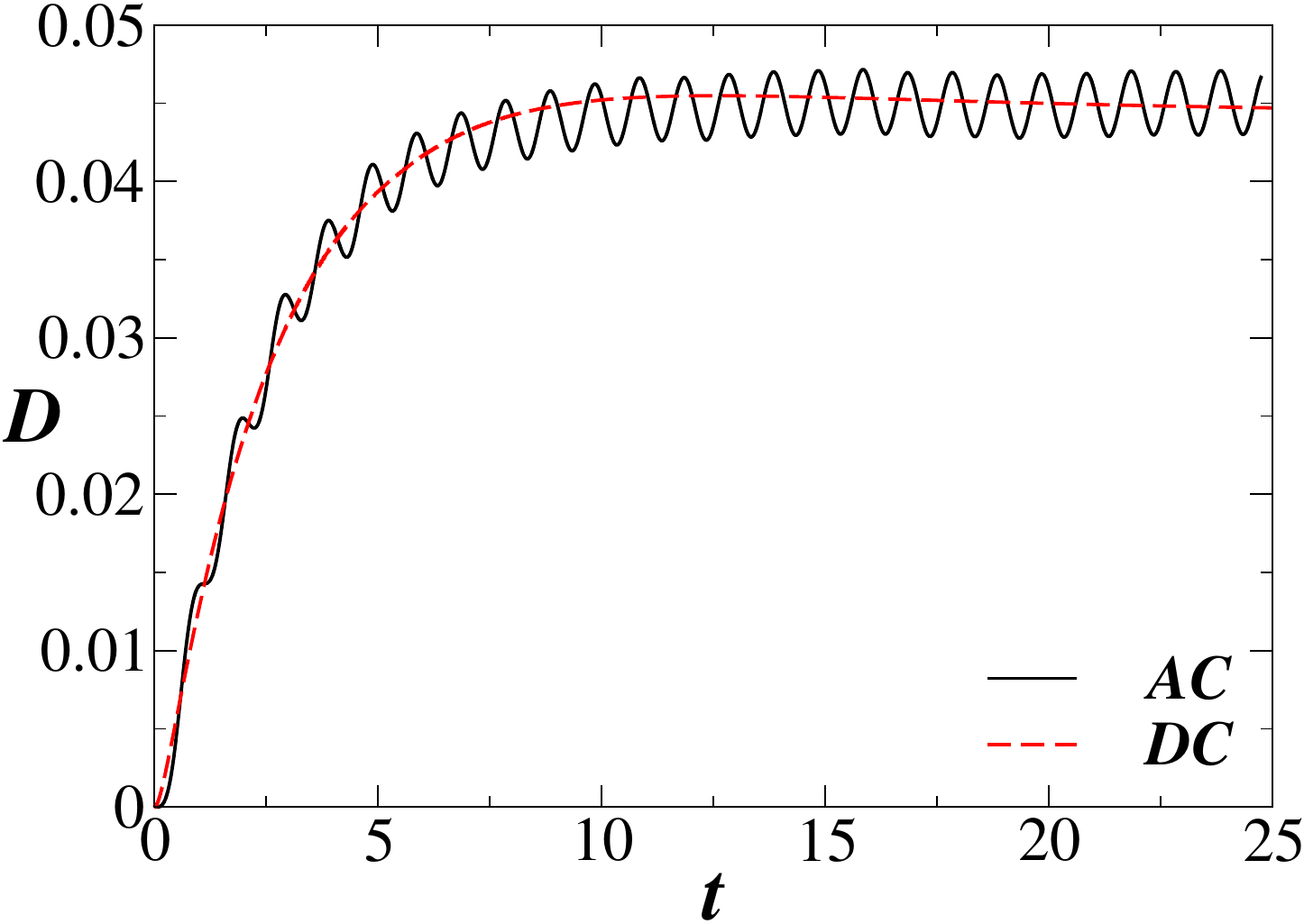}\\ 
\caption{\ks{The temporal variations of $D$ obtained from our axisymmetric simulations for $Re =1$: (A,C) $O_c=0.01$ and (B,D) $O_c=10$. The panels (A,B) and (C,D) are for ($S=10$ and $K_r=6$) and ($S=10$ and $K_r=10$), respectively. The values of potential in the alternating and direct electric fields are $\psi_0=4$ and $\psi_{DC}=4/\sqrt{2}$, respectively. $T_p=2$ is considered for all the alternating electric field cases.}}
\label{esmaeeliFig8}
\end{figure}

The pioneering work of Torza {\it et al.} \cite{torza1971} provided a deformation-circulation map in the electrical conductivity ratio $(K_r)$ and the permittivity ratio $(S)$ plane, which has been used to study EHD of a droplet under the influence of an alternating electric field. This map (shown in Fig. \ref{defmap}) delineates three different regions classified in terms of droplet deformation and flow field separated by the so-called the zero-circulation line ($S=K_r$) along which the mean velocity field is zero and the zero-deformation curve ($K_r^2 + K_r + 1 - 3S =0$). In region I, droplet deforms to a prolate and oblate shapes if $2 \pi/T_p > \omega_{cr}$ and $2 \pi/T_p<\omega_{cr}$, respectively.; for $2 \pi/T_p= \omega_{cr}$, droplet remains spherical. Here, $\omega_{cr}$ is given by $\sqrt{3 S- 1-K_r -{K_r}^2} / (O_c |S-1|)$. In regions II and III, droplet deforms to a prolate shape with a flow from poles to the equator and equator to the poles, respectively. They derived an expression for the degree of deformation, $D$ for a droplet in an unbounded axisymmetric domain in the creeping flow regime under the influence of an {alternating} electric field (Eq. (\ref{EAC})). They also verified the results experimentally. They reported that $D$ can be expressed as a sum of the mean $(D_m)$ and the oscillatory $(D_o)$ parts, i.e. $D= D_m + D_o$. The mean deformation ($D_m$) was found to be the same as the steady state deformation $(D_s)$ in case of an equivalent DC electric field (Eq. (\ref{equivDC})) when $S=K_r$ (i.e. along the zero circulation line as shown in Fig. \ref{defmap}). We begin the presentation by highlighting that this is indeed the case in Fig. \ref{esmaeeliFig8}. 

\ks{Figs. \ref{esmaeeliFig8}(A,C) and (B,D) present the comparison of temporal variations of $D$ obtained from our axisymmetric simulations using alternating and direct electric fields for $O_c=0.01$ (leaky dielectric system) and $O_c=10$ (weakly conducting system). In Figs. \ref{esmaeeliFig8}(A,B), we consider a point slightly away from the $S=K_r$ line ($S=10$ and $K_r=6$; point 1 located in region II of Fig. \ref{defmap}). It can be seen that in the leaky dielectric system (Fig. \ref{esmaeeliFig8}A), the mean degree of deformation ($D_s$) obtained under the AC field is the same as the steady state deformation obtained in the case of the equivalent DC field. In contrast, in Fig. \ref{esmaeeliFig8}B (weakly conducting system), this equivalence does not exist. Figs. \ref{esmaeeliFig8}(C,D) show the variations of $D$ for  $S=K_r$ ($S=10$ and $K_r=10$; point 2 located on the $S=K_r$ line of Fig. \ref{defmap}). In these cases, it can be observed that for both the leaky dielectric and weakly conducting systems, the mean degree of deformation obtained under the AC field and the steady state deformation obtained in the case of the equivalent DC field are the same. The theoretical prediction of $D_s$ obtained using Eq. (\ref{theory}) for $S=10, K_r=6$ and $S=10,K_r=10$ are 0.012 and 0.04, respectively. From the results presented in Fig. \ref{esmaeeliFig8}, we can conclude that, in weakly conducting systems, there is no similarity between the dynamics of a droplet under alternating and direct electric fields when $S \ne K_r$. This has also been observed experimentally \cite{torza1971}. On the other hand, as discussed in the introduction, the similarity exists in the leaky dielectric systems irrespective of the combination of $S$ and $K_r$ \cite{esmaeeli2018electrohydrodynamics,behjatian2013electrohydrodynamics,yang2017electrohydrodynamic}. Thus, in the current work, we investigate the deformation of a droplet subjected to an AC electric field and the equivalent DC electric field for weakly conducting systems.}

\begin{figure}
\centering
\hspace{0.5cm}{(A)} \\
\includegraphics[width=0.3\textwidth]{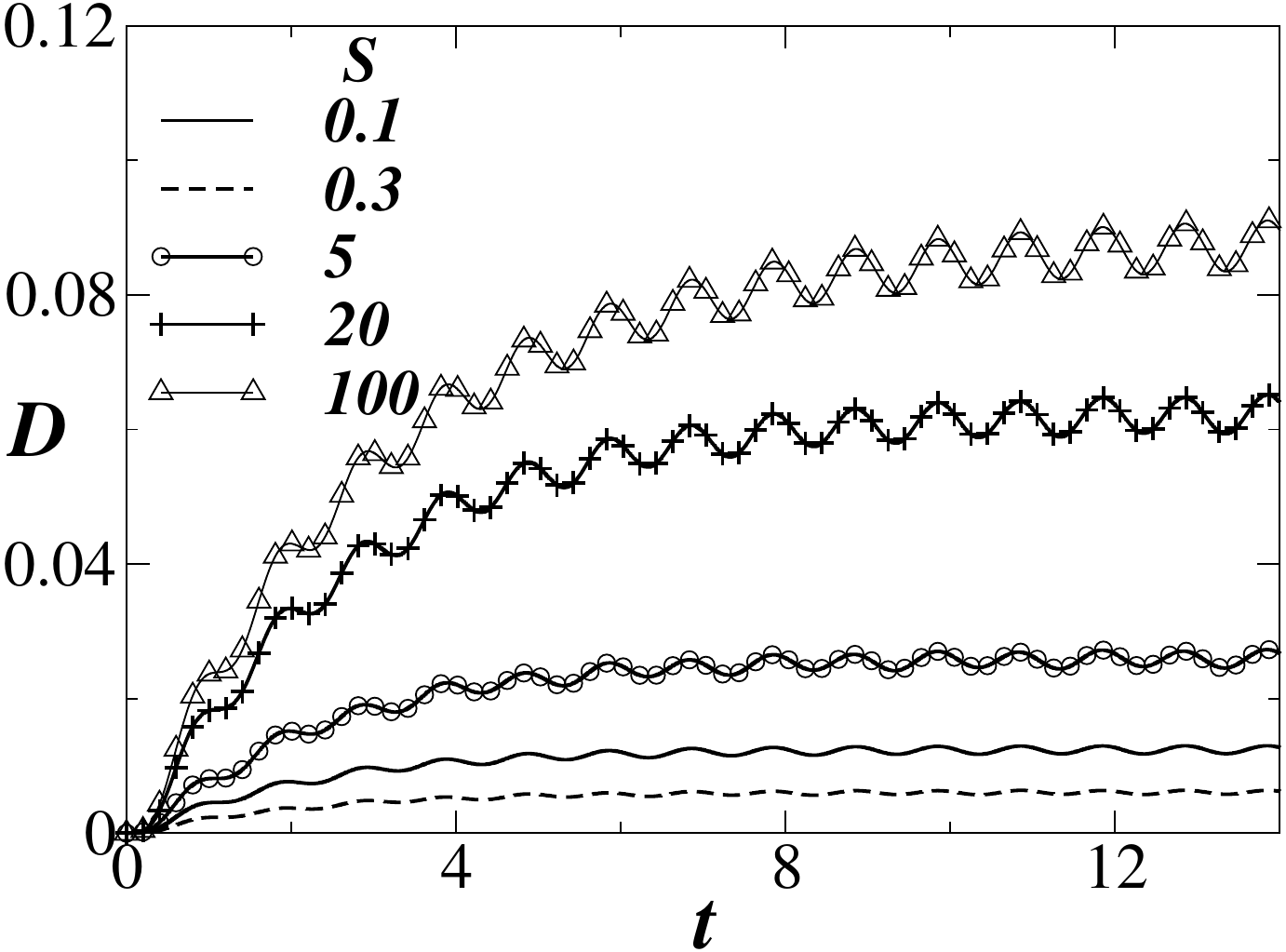} \\
\hspace{0.5cm}{(B)} \\
 \includegraphics[width=0.3\textwidth]{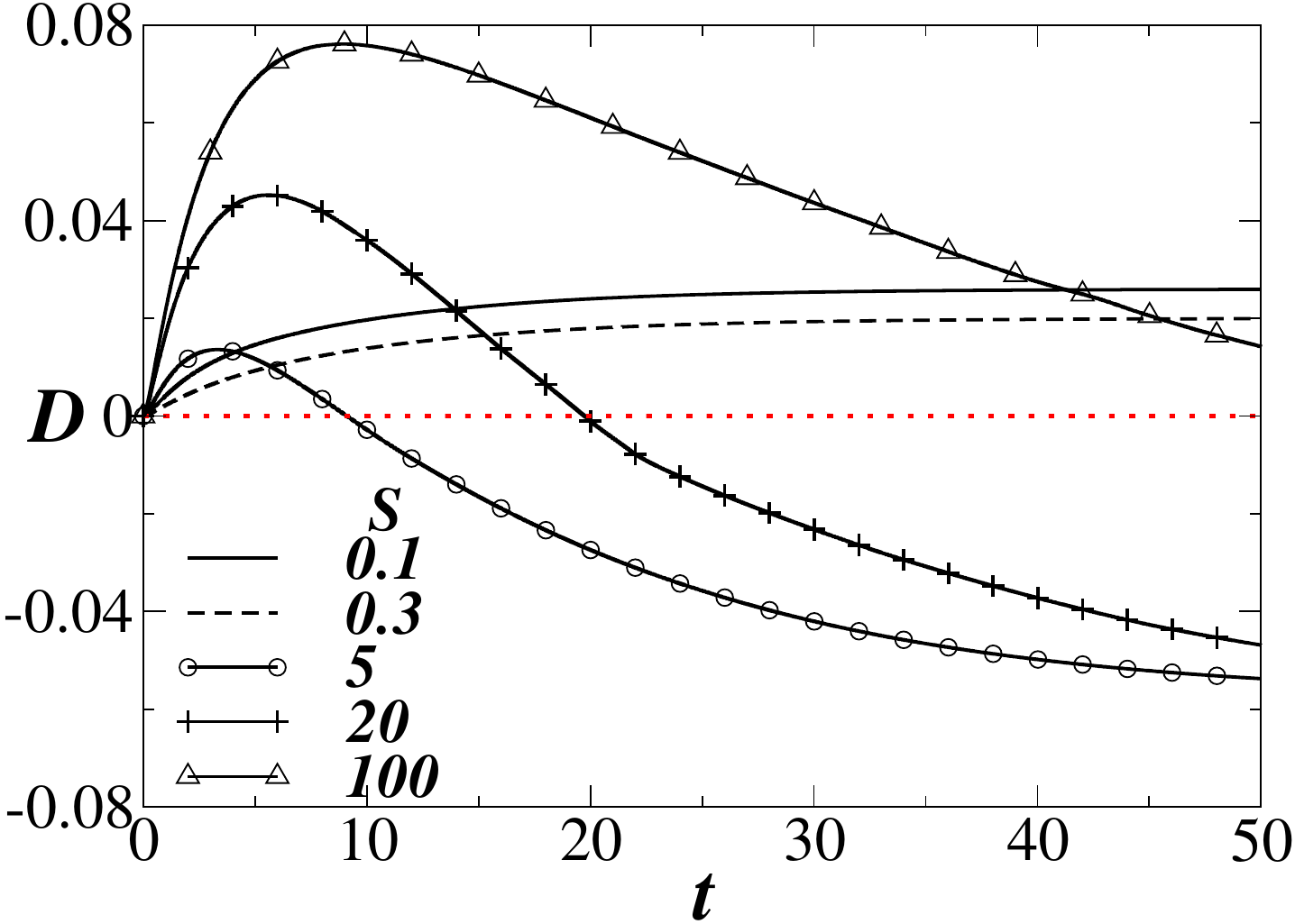} \\
\hspace{0.5cm}{(C)} \\
 \includegraphics[width=0.3\textwidth]{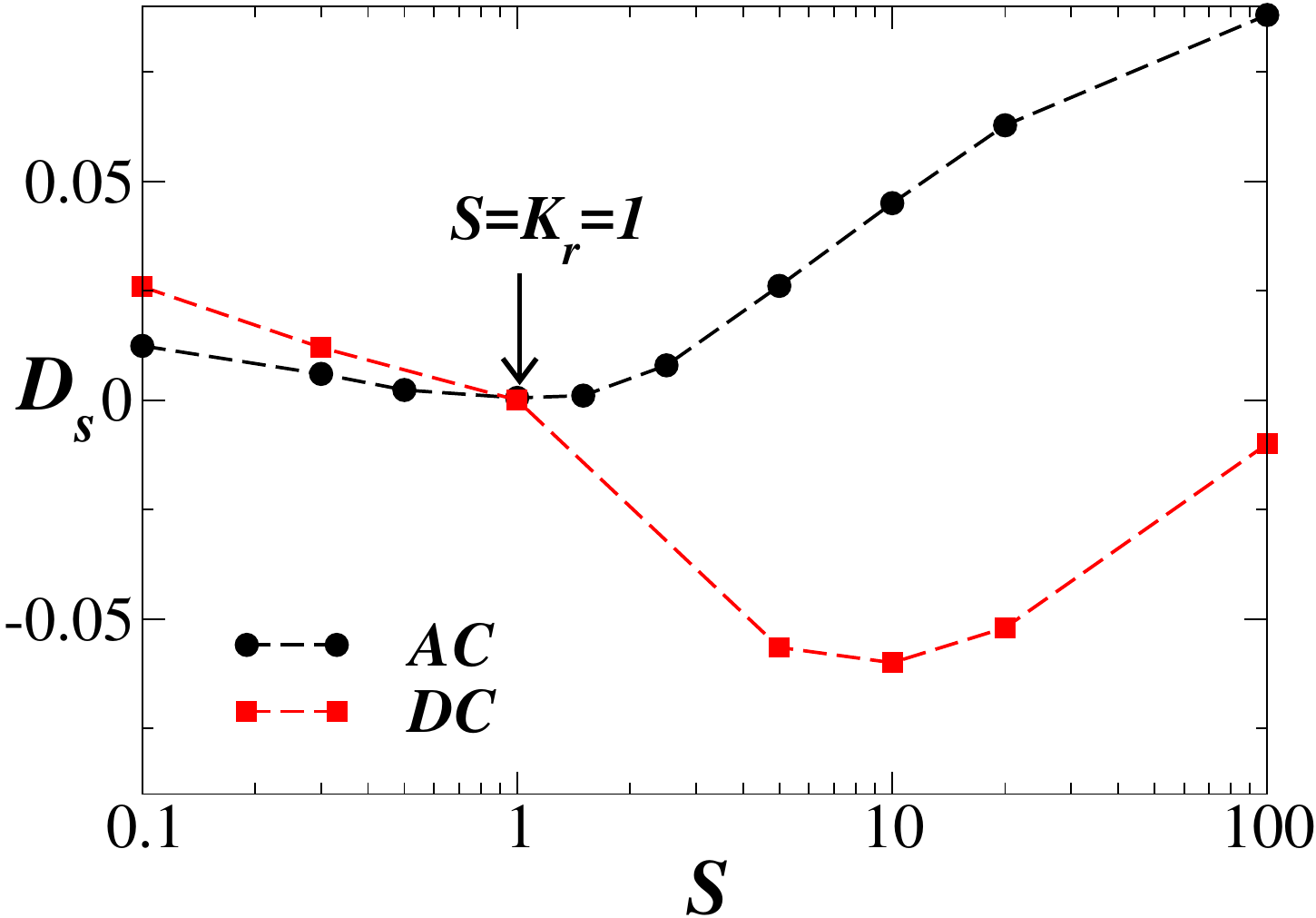}
\caption{The temporal variations of $D$ for different values of $S$ for $K_r=1$. (A) Alternating electric field ($E_0=0.5$, $T_p=2$) and (B) direct electric field ($E_{DC}=E_0/\sqrt{2}$). (C) The variation of steady mean degree of deformation, $D_s$ of the droplet with $S$ for the {alternating and direct} electric fields. The remaining parameter values are $Re =1$ and $O_c=10$. }
\label{fig8}
\end{figure}

We perform a parametric study by varying the permittivity ratio, $S$ for $K_r=1$ and the electrical conductivity ratio, $K_r$ for $S=10$. The remaining parameters are $Re =1$ and $O_c=10$. In Figs. \ref{fig8}A and B, the temporal variations of $D$ for different values of $S$ for the {alternating} electric field with $E_0=0.5$ and $T_p=2$ (which corresponds to a dimensional value of the frequency equal to 60 Hz), and the equivalent direct electric field with $E_{DC}=E_0/\sqrt{2}$ are presented, respectively. Nine values of $S$ are considered which are associated with points 4 ($S=0.1$), 5 ($S=0.3$), 6 ($S=5$), 7 ($S=20$), 8 ($S=100$), 10 ($S=10$), 14 ($S=0.5$), 15 ($S=1.5$) and 16 ($S=2.5$) in Fig. \ref{defmap}. It can be seen in Fig. \ref{fig8}A that under the action of {alternating} electric field, the droplet becomes prolate (elongates in the direction of electric field) and oscillates about a mean value of $D$ for all values of $S$ considered. For $S=0.1$ and 0.3, the droplet slightly deforms and reaches to a prolate shape (with mean degree of deformation, $D_s$) with small amplitude shape oscillations. The steady mean degree of deformation, $D_s$ of the droplet shows a non-monotonic trend with a minimum value for $S=1$ for this set of parameters. Increasing the permittivity ratio further (i.e. $S \ge 1$) increases the value of $D_s$ (see Fig. \ref{fig8}C). The amplitude of oscillations about the mean value of $D$ also increases with increasing $S$. However, the time period of oscillations is constant for all values of $S$, which is found to be half of the time period of the applied electric field, $T_p$.  

The dynamics of the droplet under the application of the equivalent DC electric field with $E_{DC}=E_0/\sqrt{2}$ is significantly different from that observed in the case of {alternating electric} field. The temporal variations of $D$ for different values of $S$ in the DC case is shown in Fig. \ref{fig8}B. In this case, for low values of $S$ ($=0.1$ and 0.3, at points 4 and 5 in region III of Fig. \ref{defmap}), the droplet deforms to a steady prolate shape; however, the value of $D_s$ decreases with increasing the value of $S$. For points in region I of Fig. \ref{defmap}, i.e. for $S=5$, 20 and 100 in Fig. \ref{fig8}B, it can be seen that the droplet initially deforms to a prolate shape (elongation) reaches to a maximum value, followed by a contraction in the direction of electric field and eventually reaches to a steady oblate shape (Fig. \ref{fig8}C). The times taken by the droplet to reach to its maximum prolate and the final steady state increase with increasing the value of $S$. 

\begin{figure}[h]
\centering
\hspace{0.5cm}{(A)} \\
\includegraphics[width=0.3\textwidth]{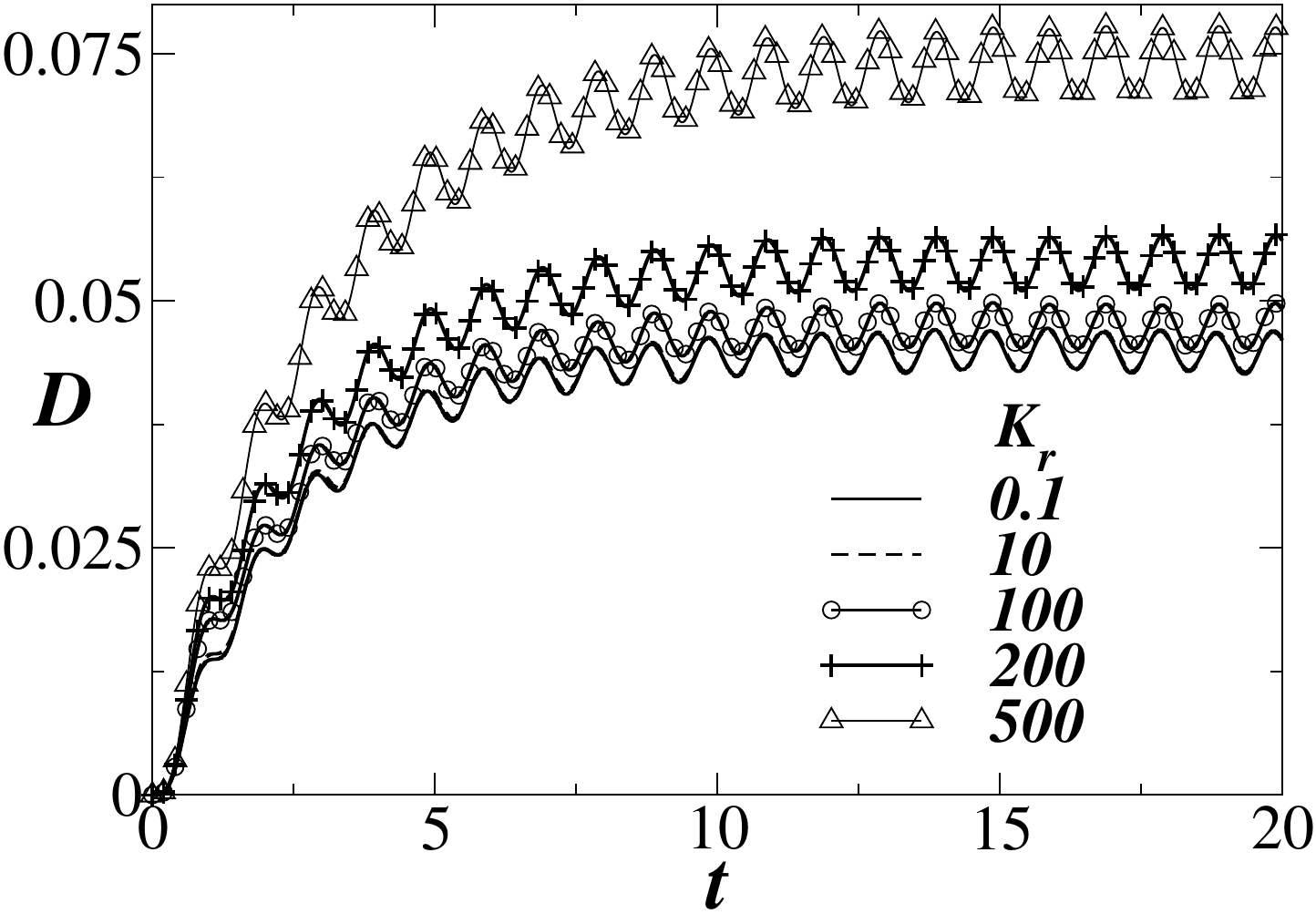} \\
\hspace{0.5cm}{(B)} \\
 \includegraphics[width=0.3\textwidth]{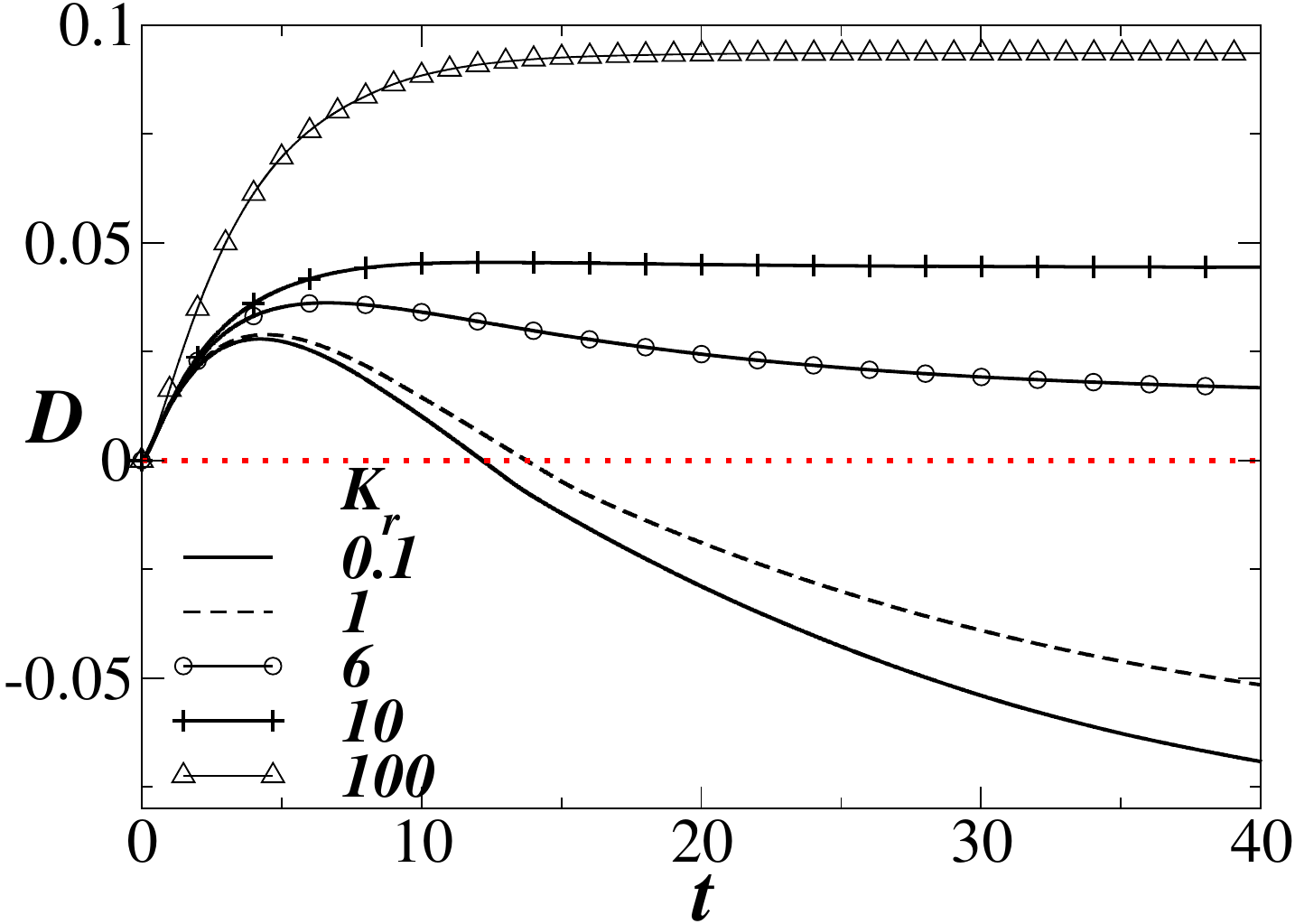} \\
\hspace{0.5cm}{(C)} \\
\includegraphics[width=0.3\textwidth]{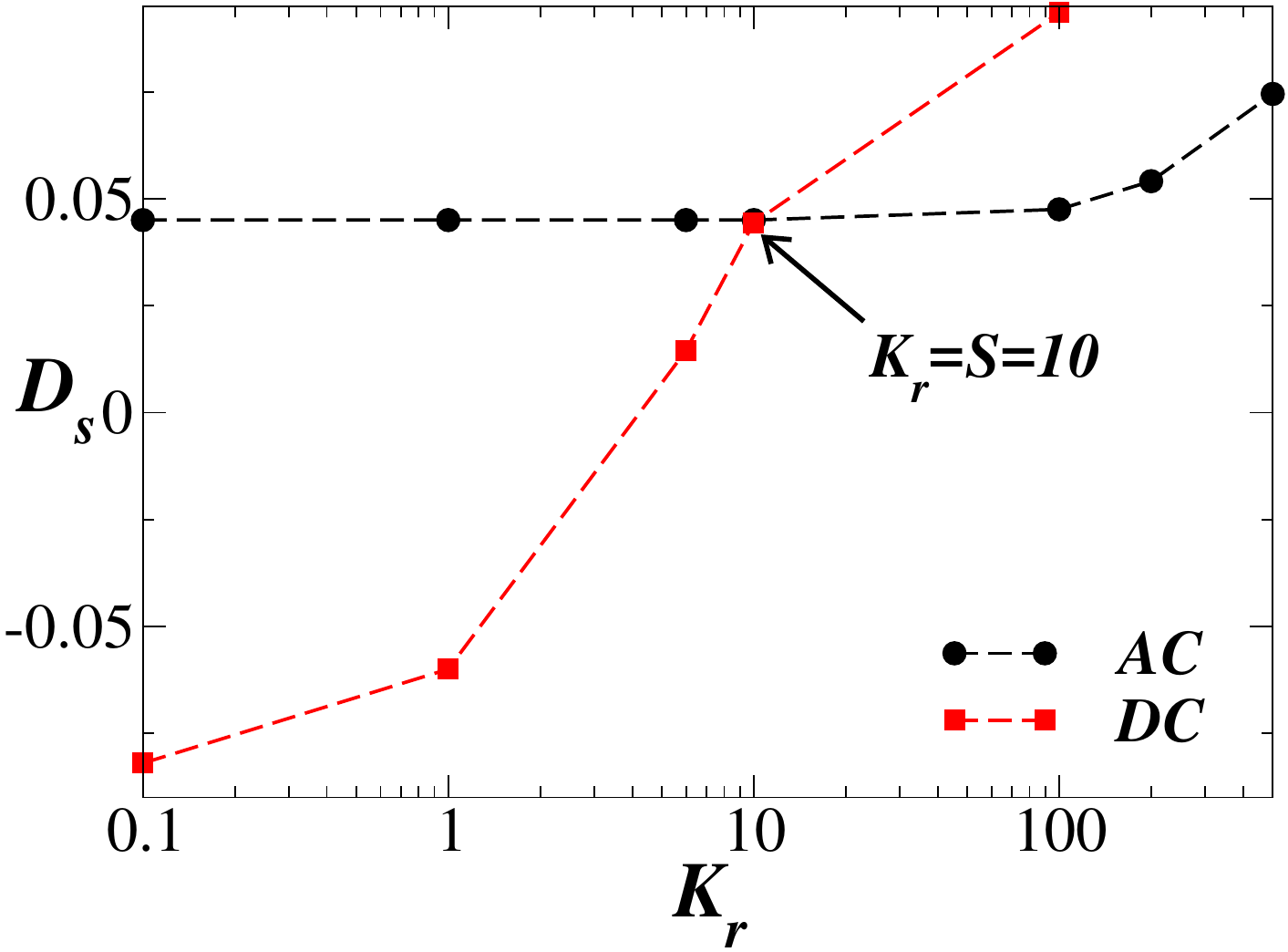}
\caption{The temporal variations of $D$ for different values of $K_r$ for $S=10$. (A) Alternating electric field ($E_0=0.5$, $T_p=2$) and (B) direct electric field ($E_{DC}=E_0/\sqrt{2}$). (C) The variation of steady mean degree of deformation, $D_s$ of the droplet with $K_r$ for the {alternating} and {direct} electric fields. The remaining parameter values are $Re =1$ and $O_c=10$.}
\label{fig9}
\end{figure}

The temporal variations of $D$ for $K_r=0.1$ (point 9), 10  (point 2), 100 (point 11), 200 (point 12) and 500 (point 13) under the action of the {alternating} electric field with $E_0=0.5$ and $T_p=2$ are presented in Fig. \ref{fig9}A. For the range of $K_r$ values considered, the droplet deforms to a prolate shape and exhibits periodic oscillations with a time period half of that of the applied electric field. It can be observed that below the $S=K_r$ line (the zero-circulation line) in Fig. \ref{defmap}, increasing $K_r$ has a negligible effect on the deformation dynamics of the droplet for the set of parameters considered in the present study. Above $S=K_r$ line, increasing $K_r$ increases the steady/mean degree of deformation of the droplet, $D_s$ (Fig. \ref{fig9}C). The amplitude of oscillations about the mean value of $D$; however with a constant time period, is found to be increased with the increase in the value of $K_r$. On the other hand, under the action of an equivalent DC electric field $(E_{DC}=E_0/\sqrt{2})$, it can be seen in Fig. \ref{fig9}B that, in region I of Fig. \ref{defmap}, the droplet deforms to a steady oblate shape after deforming to an early prolate shape (see the results for $K_r=0.1$ (point 9) and $K_r=1$ (point 10)). Above the zero-deformation curve, the droplet deforms as the time progresses and reaches a steady prolate shape (see the results of $K_r=6$ (point 1),  $K_r=10$ (point 2) and $K_r=100$ (point 11)). Close inspection of Fig. \ref{fig9}B also reveals that for $K_r<S$, the droplet always reaches to an intermediate prolate shape, followed by a decrease in the degree of deformation, and finally the droplet obtains its final steady shape. The time scale for the charge relaxation from the surrounding fluid to the interface is given by $\tau = \epsilon/K$. Thus, the ratio of charge relaxation times of the inner fluid to surrounding fluid is $S/K_r$. For $S>K_r$, the free charges in the inner fluid take longer time to relax to the interface as compared to the free charges in the surrounding fluid. Thus at the early time less than the Maxwell-Wagner relaxation timescale the drop behaves similar to a perfect dielectric and is prolate. Subsequently, as the charges relax in the inner fluid, the steady state charge distribution governs the final shape of the droplet. It can be observed in Figs. \ref{fig8}C and \ref{fig9}C that the values of $D_s$ are equal in the {alternating} and {direct} electric fields when $K_r=S$, as observed by Torza {\it et al.} \cite{torza1971}. They also experimentally observed that for high permittivity and low electrical conductivity ratios, the droplet eventually obtains a prolate shape and an oblate shape under the application of {alternating} and equivalent {direct} electric fields, respectively.

\begin{figure}
\centering
\includegraphics[width=0.3\textwidth]{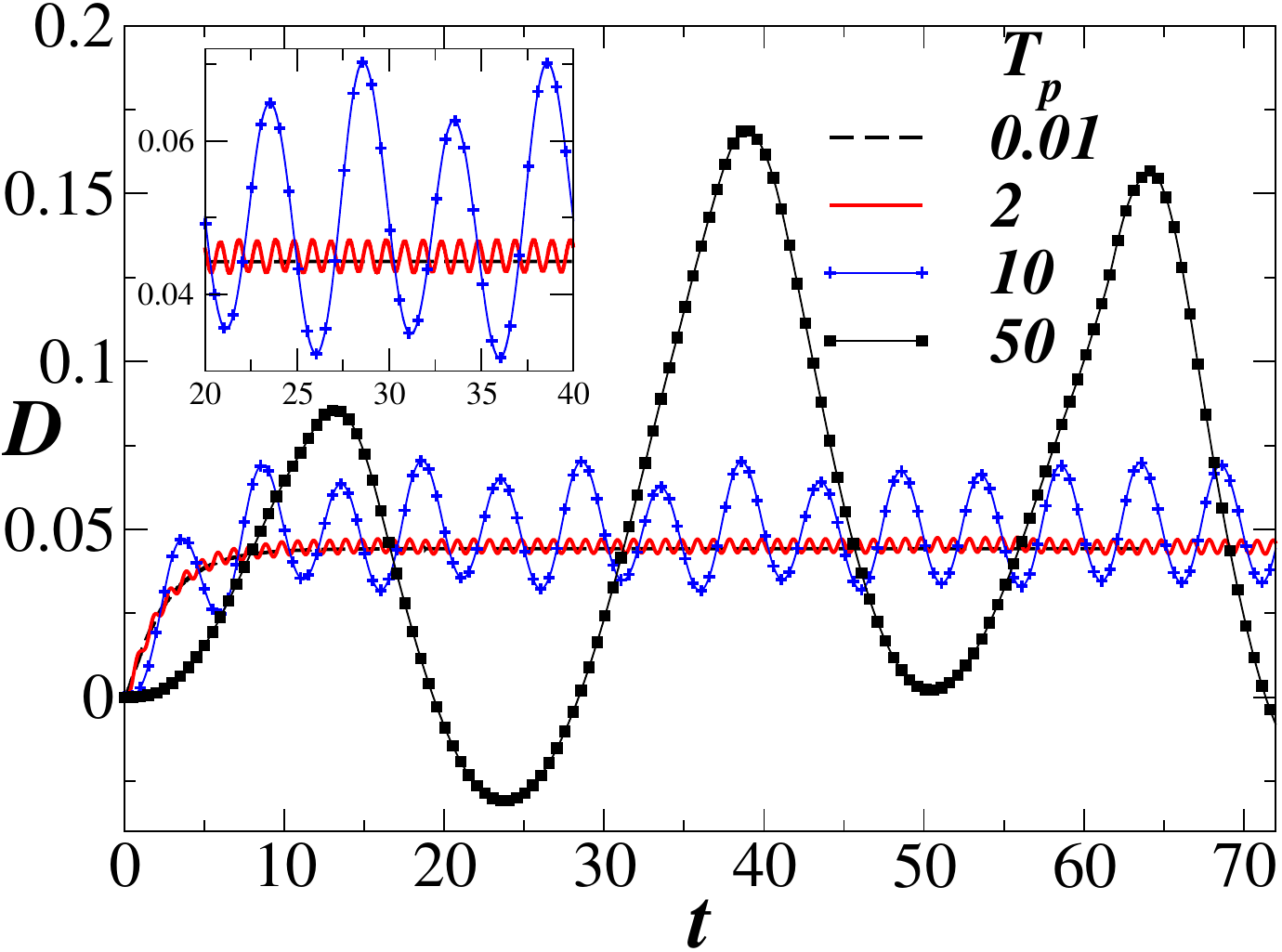}
\caption{The temporal variations of $D$ under the influence of an alternating electric field for $\psi_0=4$ with different values of $T_p$. The remaining parameter values are $Re =1$, $O_c=10$, $S=10$ and $K_r=2$. The inset represents a magnified view.}
\label{fig6}
\end{figure}

Next, in Fig. \ref{fig6}, we demonstrate the effect of the time period $(T_p)$ of the applied AC field with $\psi_0=4$ on the deformation of the droplet at point 3 ($S=10$ and $K_r=2$) in Fig. \ref{defmap}. Note that point 3 in Fig. \ref{defmap} lies in region I; away from the zero circulation $(S=K_r)$ line. The remaining parameter values are $Re =1$ and $O_c=10$. It can be seen that an initially spherical droplet undergoes periodic shape oscillations under the application of AC electric field. Increasing the value of $T_p$ increases the amplitude and time period of oscillations of the droplet. We observe that for low values of $T_p$, the droplet deforms to a prolate shape $(D>0)$ and undergoes shape oscillations about a mean degree of deformation, $D_s$ (see $T_p \le 10$ in Fig. \ref{fig6}). For  a high value of $T_p$ (see for instance, $T_p=50$ in Fig. \ref{fig6}), the droplet undergoes periodic oscillations, but during the oscillations it deforms between a slight oblate shape to a large prolate shape. The shape oscillations of the droplet become complex for high values of $T_p$ (see $T_p \ge 10)$. Close inspection of Fig. \ref{fig6} also reveals that the time period of shape oscillations of the droplet is about $T_p/2$ as also observed in earlier studies\cite{esmaeeli2018electrohydrodynamics}.

\begin{figure}
\centering
\includegraphics[width=0.3\textwidth]{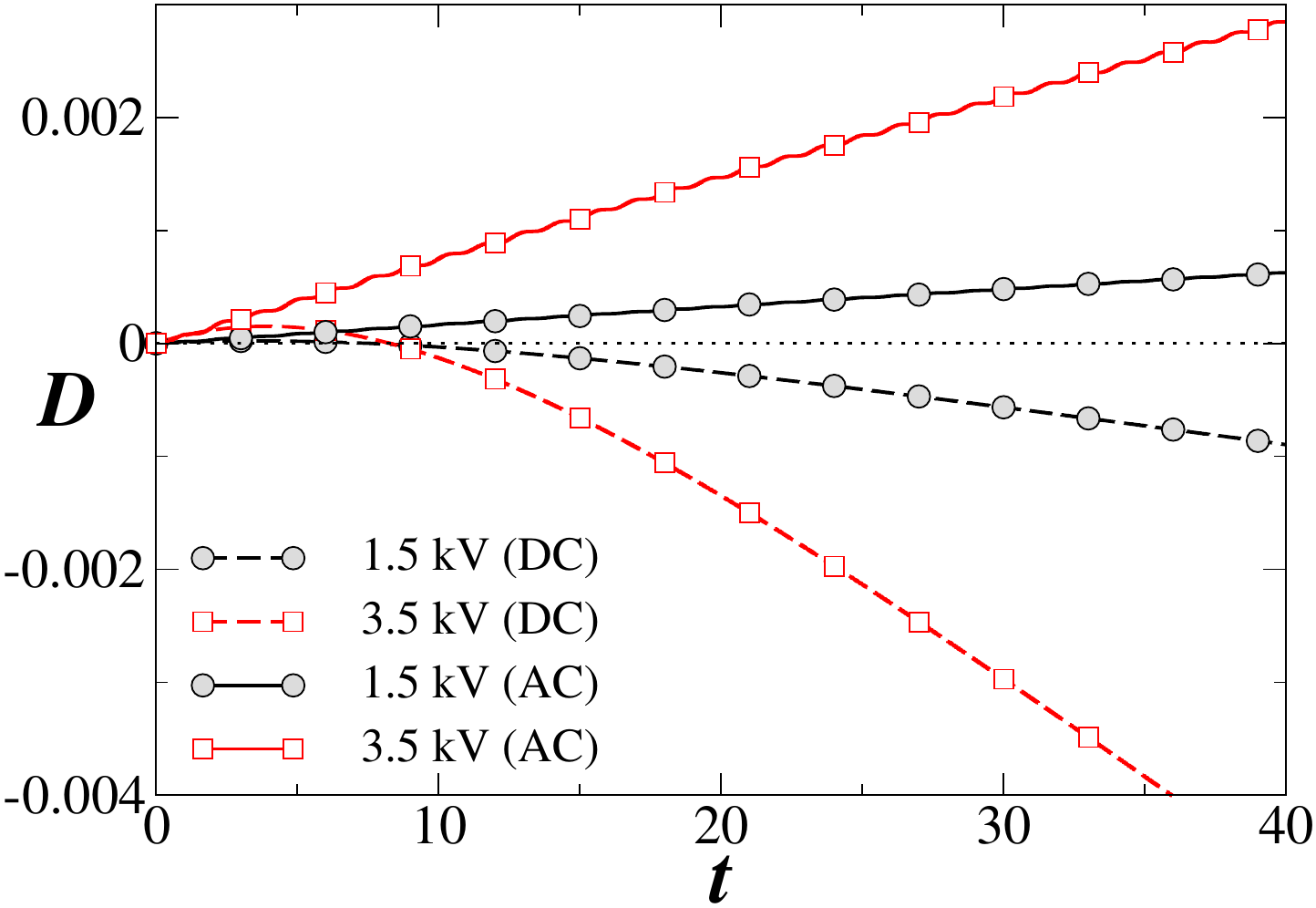} 
\caption{The temporal variations of $D$ for the parameters considered by Torza {\it et al.} \cite{torza1971}. The dashed line and the solid line represent the {results of} the {direct} and {alternating electric fields}, respectively. The dotted line corresponds to $D=0$ (spherical shape).}
\label{Torza_fig}
\end{figure}

Finally, we have performed simulations for two sets of parameters (see system 16 in Table 1 of Torza {\it et al.} \cite{torza1971}) for the {alternating} and {direct electric fields}. The droplet and the surrounding medium are silicone oil and castor oil, respectively.  The densities of both the fluids are equal to 980 kg/m$^3$. The values of the viscosity of the silicone oil and castor oil are 54 P and 65 P (nearly the same). Thus, we assume that the density and the viscosity ratios to be 1 in our simulations. The interfacial tension for this pair of fluids is $5.5 \times 10^{-3}$ N/m. The initial radius of the droplet is 0.6 mm. Two values of the DC voltage, namely, 1.5 kV and 3.5 kV are considered. The values of the associated dimensionless numbers used in our simulations are $K_r = 0.03$, $S=0.44$, $Re= 8.84 \times 10^{-3}$ and $O_c=8.9$. The dimensionless time period of the applied electric field, $T_p$ is 2.66, which corresponds to a dimensional frequency of 60 Hz as taken by Torza {\it et al.} \cite{torza1971}. The temporal variations of $D$ for the {alternating} and {direct electric fields} are shown in Figure \ref{Torza_fig}. As the times at which the shapes of the droplet are shown have not been given in Ref. \cite{torza1971}, it is not possible to compare the droplet shapes directly. Nevertheless, it can be seen in Figure \ref{Torza_fig} that the droplet elongates along and orthogonal to the direction of the applied {alternating} and {direct electric} fields, respectively. It can also be seen that the deformation increases with the increase in the applied electric forcing. This deformation behaviour can be clearly seen Figure 7 of Ref. \cite{torza1971}.

 \section*{5~~Concluding remarks}
\label{sec:conc}
\vspace{-4mm}
The electrohydrodynamics of an initially spherical droplet under the influence of an external {alternating} electric field is investigated via axisymmetric numerical simulations. \ks{In order to isolate the effect of the electric field, the system is considered to be neutrally buoyant.  The dynamic viscosity of the droplet and the surrounding medium are also assumed to be the same. A charge-conservative volume-of-fluid (VoF) based finite volume flow solver is used. The governing equations are solved without making the leaky dielectric assumption. The present work is motivated from the results reported in the earlier studies on a droplet suspended in a surrounding medium and subjected to an alternating electric field using the leaky dielectric assumption. These studies show shape oscillations about the steady-state deformation under an equivalent root mean squared direct electric field, irrespective of the electrical conductivity ratio ($K_r$) and permittivity ratio ($S$), but experimentally, this can be observed only when $K_r = S$ for a general class of fluids where the leaky dielectric assumption may not be valid \cite{torza1971}. Our simulations using weakly conducting media show that the equivalence between alternating and direct electric fields does not hold for $K_r \ne S$, thereby confirming the experimental behaviour reported by Torza {\it et al.} \cite{torza1971}.} A parametric study is conducted by varying the time period of the applied {alternating} electric field, the permittivity and the electrical conductivity ratios. Our results have revealed that for $K_r < S$, under the application of the DC electric field, the droplet deforms to a steady oblate shape after deforming to an early prolate shape. Above the $K_r=S$ line, the droplet continues to deform monotonically and attains a steady prolate shape. On the other hand, in the case of the {alternating} electric field, the droplet oscillates about a prolate shape. \ks{It is also observed that while increasing $K_r$ has a negligible effect on the deformation dynamics of the droplet below the zero-circulation line $(S=K_r)$, it enhances the deformation of the droplet above the $S=K_r$ line for both alternating and direct electric fields. The findings from current numerical simulations may help in explaining the complex behaviour of droplets in a multitude of applications ranging from digital microfluidics to medical technology.}

\vspace{-3mm}
\section*{Conflict of Interest Statement}
\vspace{-4mm}
The authors have declared no conflict of interest.
\vspace{-5mm}
\section*{Acknowledgment}
\vspace{-4mm}
KS thanks the Science and Engineering Research Board, India for providing financial support through the grant number, MTR/2017/000029. SC and MKT acknowledges the support of DST through the J. C. Bose National Fellowship and DST/INSPIRE/04/2015/000449, respectively. The valuable suggestions from anonymous reviewers are gratefully acknowledged.


\end{document}